\documentclass[nodate,preprintnumbers,amsmath,amssymb,aps,prd]{revtex4-1}
\usepackage{graphicx}
\usepackage[figurename=Fig.]{caption}

\def\qt{\ln q^{\frac{1}{3}}}

\begin{document}


\title{Intrinsic time geometrodynamics: explicit examples}



\author{Huei-Chen Lin}
\email[l28981018@mail.ncku.edu.tw]{}
\affiliation{Department of Physics, National Cheng Kung University, Tainan 701, Taiwan}
\author{Chopin Soo}
\email[cpsoo@mail.ncku.edu.tw]{}
\affiliation{Department of Physics, National Cheng Kung University, Tainan 701, Taiwan}



\begin{abstract}
 Intrinsic time quantum geometrodynamics resolved `the problem of time' and bridged the deep divide between quantum mechanics and canonical quantum gravity
 with a Schrodinger equation which describes evolution in intrinsic time variable. In this formalism, Einstein's general relativity is a particular realization of a wider class of theories.
  Explicit classical black hole and cosmological solutions and the motion of test particles are  derived and analyzed in this work
 in the context of constant three-curvature solutions in intrinsic time geometrodynamics; and we exemplify how this formalism yields results which agree with the predictions of Einstein's theory.
 \\
 \\
{{\bf  Note:} This article is an extension of\\ {\it Intrinsic time geometrodynamics: explicit examples},\\ Huei-Chen Lin and Chopin Soo, Chin. J. Phys. {\bf 53}, 110102 (2015),\\
published in Chinese Journal of Physics (Volume 53, Number 6 (November 2015)), {\rm Special Issue} {\it On the occasion of 100 years since the birth of Einstein's General Relativity}}

\end{abstract}

\maketitle
\section{Introduction to Intrinsic Time Geometrodynamics}

A framework for geometrodynamics without the paradigm of space-time covariance has been advocated in Refs.\cite{Soo_Hoi-Lai,  Niall, ITQG, NCR}.
With a Schrodinger equation for quantum geometrodynamics which describes first-order evolution in intrinsic time, it resolved `the problem of time' and bridged
 the deep divide between quantum mechanics and conventional canonical formulations of quantum gravity.
In Horava-Lifshitz gravity\cite{Horava2009},  the deep conflict between gravity as a unitary field theory and space-time 4-covariance is overcome by retaining only spatial covariance in the theory; and
power-counting renormalizability is achieved by supplementing the potential of Einstein's theory with higher spatial curvature terms which explicitly break 4-covariance. In this work, classical black hole and cosmological solutions
and the motion of point particles are derived and discussed from the point of view of intrinsic time geometrodynamics, and we demonstrate how this formalism can yield results which agree with the predictions of Einstein's theory.

To recapitulate the theory, we start with the Arnowitt-Misner-Deser (ADM) decomposition
$ds^2=-N^2dt^2+q_{ij}\left(dx^i+N^idt\right)\left(dx^j+N^jdt\right)$.
The canonical action of General Relativity (GR) may be written as
\begin{equation}
S=\int{\tilde\pi}^{ij}\dot{q}_{ij}d^3xdt-\int\left(NH+N^iH_i\right)d^3xdt + {\rm boundary\,\, term},
\end{equation}
wherein the super-Hamiltonian $H=\frac{2\kappa }{\sqrt{q}}\left[G_{ijkl}{\tilde\pi}^{ij}{\tilde\pi}^{kl}+ V(q_{ij})\right]$, and
$H_i=-2q_{ij}\nabla_k{\tilde \pi}^{kj}=0$ is the super-momentum constraint which generates spatial diffeomorphisms of the variables.
The DeWitt supermetric, with deformation parameter $l$, is $G_{ijkl}=\frac{1}{2}\left(q_{ik}q_{jl}+q_{il}q_{jk}\right)-l
q_{ij}q_{kl}$. In Einstein's  GR,  $l= \frac{1}{2}$  and the potential $V=-\frac{q}{(2\kappa)^2}(R - 2\Lambda_{\it{eff}}) $.
With $\beta^2=l-\frac{1}{3}$, the super-Hamiltonian constraint factorizes in an interesting
way as
\begin{equation}
\frac{\sqrt{q}}{2\kappa}H=G_{ijkl}\tilde{\pi}^{ij}\tilde{\pi}^{kl}+V(q_{ij})=-\left(\beta{\tilde\pi}-\bar{H}\right)\left(\beta{\tilde\pi}+\bar{H}\right)=0,
\end{equation}
wherein
$\bar{H}=\sqrt{\bar{G}_{ijkl}\bar{\pi}^{ij}\bar{\pi}^{kl}+V(q_{ij})}=\sqrt{\frac{1}{2}\left[\bar{q}_{ik}\bar{q}_{jl}+\bar{q}_{il}\bar{q}_{jk}\right]\bar{\pi}^{ij}\bar{\pi}^{kl}+V(q_{ij})}$.
In the symplectic potential $\int\tilde{\pi}^{ij}\delta
q_{ij}=\int\bar{\pi}^{ij}\delta\bar{q}_{ij}+{\tilde\pi}\delta\ln
q^{\frac{1}{3}}$, so clean separation of the conjugate pair, $(\ln q^{\frac{1}{3}}, {\tilde\pi})$, consisting of (one-third of) the logarithm of the determinant of the spatial metric and the trace of the momentum, from $({\bar q}_{ij}, {\bar \pi}^{ij})$, the unimodular part of the spatial metric with traceless conjugate momentum, allows a deparametrization of the theory wherein $\ln q^{\frac{1}{3}}$  plays the role of the intrinsic time variable for $\beta^2 >0$. In the quantum context, with $\hat{\tilde\pi} =\frac{\hbar}{i}\frac{\delta}{\delta \ln q^{\frac{1}{3}}}$  in the metric representation, the Hamiltonian constraint implies the Schr\"{o}dinger equation $(\beta\hat{\tilde\pi }+ {\bar H})\Psi =0$,
 which is a Schwinger-Tomonaga equation with multi-fingered time $\ln q^{\frac{1}{3}}(x)$ .

Hodge decomposition for compact manifolds without boundary yields
$\delta \ln q^{\frac{1}{3}} =\delta{T}+\nabla_i\delta{Y}^i$, wherein $\delta T =\frac{2}{3}\delta \ln V_{spatial}$ (which is proportional to the logarithmic change in the spatial volume) is a three-dimensional diffeomorphism invariant (3dDI) quantity which serves as the global (spatially independent)  intrinsic time interval; whereas $\nabla_i\delta{Y}^i$ can be gauged away since the Lie derivative ${\mathcal L}_{\delta{{\overrightarrow N}}}\ln q^{\frac{1}{3}} = \frac{2}{3}\nabla_i{\delta N^i}$.  With respect to time-development in $\delta T$,  the 3dDI physical Hamiltonian is ${H}_{\rm Phys}:=\int\frac{{\bar H}(x)}{\beta}d^3x$, with fundamental 3dDI Schr\"{o}dinger equation, $i\hbar \frac{\delta \Psi}{\delta T} = {H}_{\rm Phys}\Psi$.  In the classical context,  it has been demonstrated, through the Hamilton-Jacobi and Hamilton equations, in Refs.\cite{Soo_Hoi-Lai,Niall} that the resultant classical spacetime that is produced by this theory has an {\it emergent} ADM lapse function which is precisely
\begin{equation}\label{lapse}N =\frac{\sqrt{q}(\partial_t \ln q^{1/3} -\frac{2}{3}\nabla_iN^i)}{4\beta\kappa {\bar H}}.
\end{equation}
In conventional canonical formulation of Einstein's GR, the EOM  $\frac{dq_{ij}}{dt} = \big{\{} q_{ij}, \int {N} H + N^iH_i\big{\}}_{P.B.}
= \frac{4{N}\kappa}{\sqrt q}{G_{ijkl}}{\tilde\pi}^{kl}  + {\cal L}_{\vec{N}} q_{ij}$,
relates the extrinsic curvature to ${\tilde\pi}^{ij}$ by
$K_{ij} := \frac{1}{2N}(\frac{dq_{ij}}{dt} - {\cal L}_{\vec{N}}q_{ij}) = \frac{2\kappa}{\sqrt q}G_{ijkl}{\tilde\pi}^{kl}$. Taking the trace gives
$\frac{1}{3}K=\frac{1}{2N}\big(\partial_t{\qt} -  \frac{2}{3}{\nabla_i N^i}\big)= \frac{2\kappa\beta}{\sqrt q}{\bar H}$,
wherein the constraint  $(\beta{\tilde\pi} +{\bar H})=0$ has been used to arrive at the last step. Thus it is noteworthy that {\it Einstein's theory yields an a posteriori lapse function $N$}  which agrees physically with the result \eqref{lapse} of the classical spacetime produced by ${H}_{\rm Phys}$ in the formalism of intrinsic time geometrodynamics. Further details and remarks on the formalism can be found in the appendix.

From the perspective of intrinsic time geometrodynamics and the paradigm shift to 3dDI, Einstein's GR (with its corresponding $V$ and $\beta$ ) is a particular realization of a wider class of theories.
Requirement of a real physical Hamiltonian density compatible with spatial diffeomorphism symmetry suggests supplementing the kinetic term with a quadratic
form, i.e.
\begin{equation}\label{Hbar}
\bar{H}=\sqrt{\bar{G}_{ijkl}\bar{\pi}_{ij}\bar{\pi}_{kl}+\left[\frac{1}{2}(q_{ik}q_{jl}+q_{jk}q_{il})+\gamma
q_{ij}q_{kl}\right]\frac{\delta W}{\delta q_{ij}}\frac{\delta
W}{\delta q_{kl}}}.
\end{equation}
$\bar{H}$ is then real if $\gamma>-\frac{1}{3}$.
Dependent only on 3-geometry, $W$ is the sum of a
Chern-Simons action of the spatial affine connection and the spatial
Einstein-Hilbert action with the cosmological constant i.e.
\begin{equation}\label{alpha_Lambda}
W=\int\left[\sqrt{q}\left(b
{\cal R}-\Lambda\right)+g\tilde{\epsilon}^{ikj}\left(\Gamma^l_{im}\partial_j\Gamma^m_{kl}+\frac{2}{3}
\Gamma^l_{im}\Gamma^m_{jn}\Gamma^n_{kl}\right)\right].
\end{equation}
A slight generalization  is to replace $\frac{\delta W}{\delta { q}_{ij}}$ in the positive semidefinite quadratic form in ${\bar H}$ with ${\sqrt q}(\Lambda' q^{ij}+ b' {\cal R}q^{ij} + c'{\bar{\cal R}}^{ij}+ g'C^{ij})$, which is the most general symmetric second rank tensor (density) containing up to third derivatives of the spatial metric\cite{Soo_Hoi-Lai}. Quantum considerations which prompt further improvements in the precise expression of the Hamiltonian are detailed in Refs.\cite{ITQG,NCR}.

\section{Explicit solutions}

In addition to the spatial Ricci scalar and cosmological constant, higher curvature terms such as the traceless part of the Ricci tensor, ${\bar{\cal R}}_{ij}$, and the Cotton-York tensor, $C_{ij}$, are thus present in the potential in the generalized Hamiltonian density $\bar H/\beta$. But these vanish identically for solutions with constant spatial curvature slicings, which implies the physical content of well-known solutions of Einstein's theory can be captured by ${H}_{\rm Phys}$ of intrinsic time geometrodynamics in this setting.
\subsection{Constant spatial curvature slicings}

 We consider constant 3-curvature slicings with
$t$-independent lapse and the following shift vector for simplicity to obtain {\it exact solutions of the full theory}.
As we shall show explicitly, these will include the Robertson-Walker and Painleve-Gullstrand form of Schwarzschild-deSitter solutions.
The metric may then be expressed rather generically as
\begin{equation}\label{RW_metric}
ds^2=-N(r)^2dt^2+a^2(t)\left[\left(\frac{dr}{\sqrt{1-kr^2}}+n(r)dt\right)^2+r^2\left(d\theta^2+\sin^2\theta
d\phi\right)\right], \end{equation} with the constant spatial scalar curvature
 ${\cal R} =\frac{6k}{a^2}$, and $k=0, \pm 1$ determines the topology of the slicings.
 For the above constant 3-curvature slicings, the super-Hamiltonian constraint simplifies
 to $H \propto K_{ij}K^{ij}+\frac{l}{1-3l}K^2+\lambda =0$, with $\lambda :=2\Lambda_{eff}-{\cal R}+\frac{{\cal R}^2}{8\Lambda_{eff}}=
2\Lambda_{eff} -\frac{6k}{a^2}+\frac{9k^2}{2\Lambda_{eff}a^4}$. And the extrinsic curvature (for t-independent $a$) can thus be determined
as
\begin{equation}\label{Keqn} K_{ij}=\frac{1}{2N}\left(\partial_t
q_{ij}-\nabla_i N_j-\nabla_j N_i\right)=\left(\begin{array}{ccc}
-\frac{a^2(\partial_rn)}{N\sqrt{1-kr^2}} &0 &0\cr
0&-\frac{ra^2n\sqrt{1-kr^2}}{N} &0\cr
0&0&-\frac{ra^2n\sin^2\theta\sqrt{1-kr^2}}{N}\end{array}\right).\end{equation}
The super-Hamiltonian and super-momentum constraints reduce to the restrictions
\begin{equation}\label{non-linear third-order ODE}
\frac{4\partial_r{\cal G}}{(1-2l)(6{\cal G}^2-{\cal G}^3-12{\cal G})+16(1-l)}=\frac{1}{r}
,\end{equation} with ${\cal G}=\frac{2r(\partial_rn)}{n}$, and
$N=\left\{\frac{\left(1-kr^2\right)\left[2(1-l)n^2+4l
rn(\partial_rn)+r^2(1-2l)(\partial_rn)^2\right]}{r^2(3l-1)\lambda}\right\}^{\frac{1}{2}}$.
From solutions ${\cal G}(r)$ of Eq.\eqref{non-linear third-order ODE},
the lapse and the shift can be determined. However, it is hard
to solve Eq.\eqref{non-linear third-order ODE} explicitly, except in the
$l=\frac{1}{2}$ limit of Einstein's theory. The implication is that
exact solutions of deformed GR  expressed in terms of $r$ is
complicated. Remarkably, it is possible to solve for the metric in terms of the
dynamical variable $\pi$ for arbitrary $l$. We discuss how this can be carried out below.

\subsection{Solution of  the constraints}
The momentum ${\tilde\pi}_{ij}\propto {\sqrt q}\,p_{ij}$ for the t-dependent constant 3-curvature
metric  of the form compatible with \eqref{Keqn} is
$p^i_{\;\;j}=\left(\begin{array}{ccc}\lambda_1 &0&0\cr 0
&\lambda_2 &0\cr 0&0 &\lambda_2\end{array}\right)$; correspondingly
\begin{align}
\lambda_1&=\frac{1}{3}\left(p+\sqrt{6p_{ij}p^{ij}-2p^2}\right), \cr
\lambda_2&=\frac{1}{3}\left(p-\frac{1}{2}\sqrt{6p_{ij}p^{ij}-2p^2}\right),
\end{align}
with the decomposition
$p^i_{\,j}=\bar{p}^i_{\,j}+\frac{1}{3}\delta^i_{\,j}p$. The traceless part
 $\bar{p}^i_{\,j}$  can thus be expressed as
\begin{equation}\label{constraint_included}\bar{p}^i_{\,j}
=\left(\begin{array}{ccc}2\sqrt{\frac{1}{6}\left(l-\frac{1}{3}\right)p^2-
\frac{1}{6}\lambda} &0&0 \cr
0&-\sqrt{\frac{1}{6}\left(l-\frac{1}{3}\right)p^2-\frac{1}{6}\lambda}&0\cr
0&0&-\sqrt{\frac{1}{6}\left(l-\frac{1}{3}\right)p^2-\frac{1}{6}\lambda}.\end{array}\right).\end{equation}
In Eq.\eqref{constraint_included} we have used the relation
$p_{ij}=K_{ij}+\frac{l}{1-3l}q_{ij}K$ and the
super-Hamiltonian constraint $H=0$ which relates $\lambda$ to $K_{ij}$ . And the super-momentum constraint
now reduces to the single requirement
\begin{equation}\label{relation_r_pi_ODE}\frac{1}{3}\partial_rp+\frac{6}{r}\sqrt{\frac{1}{6}\left(l-\frac{1}{3}\right)p^2-\frac{1}{6}\lambda}+
2\partial_r\sqrt{\frac{1}{6}\left(l-\frac{1}{3}\right)p^2-\frac{1}{6}\lambda}=0.\end{equation}
It is helpful to convert
Eq.\eqref{relation_r_pi_ODE} into
\begin{equation}\label{partial_ln}\partial_{\ln
r}p=\frac{3\sqrt{2}\left[3\lambda-(3l-1)p^2\right]}{\sqrt{2}(3l-1)p+\sqrt{(3l-1)p^2-3\lambda}}.\end{equation}
The solution reveals the relation
between $r$ and $p$ as
\begin{equation}\label{r=g(pi)}
r=\frac{\left[(3l-1)p+\sqrt{3l-1}\sqrt{(3l-1)p^2-
3\lambda}\right]^{-\frac{1}{3\sqrt{2(3l-1)}}}}{c_1\left[(3l-1)p^2-3\lambda\right]^{\frac{1}{6}}}=:g(p),
\end{equation}
wherein $c_1$ is the constant of integration. For the constant
3-curvature spacetimes of Eq.\eqref{RW_metric}, there are
two further equations for consistency:
$\bar{K}^i_{\,j}=\bar{p}^i_{\,j}$ and $K=(1-3l)p$. In
terms of $r$ and $p$ these are expressed as
\begin{equation}\label{consistency}\left\{\begin{array}{l}
\frac{\sqrt{1-kr^2}(n-r\partial_rn)}{3rN}=\sqrt{\frac{1}{6}\left(l-\frac{1}{3}\right)p^2-\frac{1}{6}\lambda},\cr
\frac{\sqrt{1-kr^2}(2n+r\partial_rn)}{rN}=(3l-1)p.
\end{array}\right.
\end{equation}
With Eq.\eqref{partial_ln}, $n$ (which is related to the shift) and
the lapse function  $N$ can be determined from Eq.\eqref{consistency} as
\begin{align}\label{lapse_shift_t_independent}
&n=\frac{\left[(3l-1)p+\sqrt{3l-1}\sqrt{(3l-1)p^2-
3\lambda}\right]^{\frac{\sqrt{2}}{3\sqrt{3l-1}}}}{\left[(3l-1)p^2-3\lambda\right]^{\frac{1}{6}}},\cr
&N=\frac{3\sqrt{2}c_1\left[(3l-1)p+\sqrt{3l-1}\sqrt{(3l-1)p^2-3\lambda}\right]^{\frac{1}{\sqrt{2(3l-1)}}}
\sqrt{1-kg(p)^2}} {\sqrt{2}(3l-1)p+\sqrt{(3l-1)p^2-3\lambda}}.
\end{align}
This completes the solution, and the last equation is a specialization of formula \eqref{lapse}. Note that while we are unable to express $p$ in terms of $r$ nicely, we are able to achieve the inversion $r =g(p)$ in \eqref{r=g(pi)} and thus the metric of \eqref{RW_metric} which now
{\it satisfies all constraints} is expressible in terms of coordinate variables $ (t, p, \theta, \phi)$ and functions $a, N(p), n(p)$ for {\it arbitrary} $l$ .

\subsection{Constant curvature slicings with t-independent scale factor $a$ }
With $a$ (hence $\lambda$) being a space-time independent constant, the generic solution for arbitrary $l$ can be written down from the formulas displayed earlier.
In the special case of $l=\frac{1}{2}$  for GR, inverting Eq.\eqref{r=g(pi)}  yields
$p=\frac{1}{(c_1r)^{\frac{3}{2}}}\left[\frac{2\left(1+\frac{3c_1^3r^3\lambda}{\sqrt{2}}\right)}{\sqrt{2\sqrt{2}+3c_1^3r^3\lambda}}\right]$.
The function $n$ and the lapse $N$ are then determined from
Eq.\eqref{lapse_shift_t_independent} to be \begin{equation}\label{1/2}
n=3c_1\sqrt{\frac{\sqrt{2}}{9c_1^3r}+\frac{\lambda}{6}r^2}; \qquad
N=3c_1\sqrt{1-kr^2}.
\end{equation}
Comparison with physics identifies the constant $c_1$ as
$c_1=\left(\frac{1}{9\sqrt{2}Gm}\right)^{\frac{1}{3}}$. Defining $dt_{\rm PG}=3c_1dt$  yields the Schwarzschild-deSitter solution as\begin{eqnarray}
ds^2=-(1-kr^2)dt_{\rm PG}^2+a^2\left[\left(\frac{dr}{\sqrt{1-kr^2}}+dt_{\rm PG}\sqrt{\frac{2Gm}{r}+\frac{\lambda}{6}r^2}\right)^2+r^2\left(d\theta^2+\sin^2\theta
d\phi^2\right)\right],\label{SdS1}\\
=-\left(1-\frac{2GM}{c^2R} -
\frac{\Lambda_{eff}}{3}R^2-\frac{3k^2R^2}{4a^4\Lambda_{eff}}\right)d\tau^2
+\frac{dR^2}{\left(1-\frac{2GM}{c^2R} -
\frac{\Lambda_{eff}}{3}R^2-\frac{3k^2R^2}{4a^4\Lambda_{eff}}\right)}+
R^2 d\Omega^2;\label{SdS2}
\end{eqnarray}
wherein $\lambda=2\Lambda_{eff}-\frac{6k}{a^2} + \frac{9k^2}{2\Lambda_{eff}a^4}= \left(\sqrt{2\Lambda_{eff}} -\frac{3k}{a^2\sqrt{2\Lambda_{eff}}} \right)^2$ , $ R:= ar$ and $M:=ma^3c^2$.
Expression, (17) is the solution written in Painleve-Gullstrand (PG) form, with constant 3-curvature slicings\cite{Lin_Soo2009}; while  the identification $d\tau = dt_{PG} -\frac{adr}{\sqrt{1-kr^2}}\frac{\sqrt{\frac{2GM}{R} + \frac{\lambda}{6}R^2}}{[1-\frac{2GM}{R} - \frac{\lambda}{6}R^2-kr^2]}$ yields
 the metric in the standard form of \eqref{SdS2}, which suffers from coordinate singularities and is a priori defined only in the region between the horizons.
 The PG form of the metric \eqref{SdS1} is free of coordinate singularities and extends the manifold beyond the horizons.
 In the case of spatially compact ($k=1$) slicing, the range of the radial coordinate is $0\leq R=ra \leq a$, and to cover a region of sufficient interest of the Schwarzschild-deSitter manifold we should choose $a\gg \sqrt{\frac{3}{\Lambda_{eff}}}$ which is the radius of the deSitter horizon (which implies $\frac{3k}{4a^4\Lambda_{eff}} \ll \frac{\Lambda_{eff}}{3}$). In fact by choosing either $k=0$, or $a^2\sqrt{\Lambda_{eff}} \mapsto \infty$  for $k=1$, the PG form \eqref{SdS1} remains valid (except at the physical singularity at $R=0$) for  the full range $R >0$ of the radial coordinate. Moreover, such a choice guarantees the physical requirement that the metric is asymptotically de Sitter with $\Lambda_{eff}$ as the value of the cosmological constant.

 \subsection{Time-dependent scale factor,  vanishing shift vector and Robertson-Walker solution}
The case with zero shift vector and t-dependent $a(t)$ in
Eq.\eqref{RW_metric} can be solved readily. For vanishing shift vector, the extrinsic curvature becomes
$K^i_{\,j}=\delta^i_{\,j}\frac{\partial_ta}{Na}$. Since
$\bar{K}^i_{\,j}=\bar{p}^i_{\,j}=0$, we obtain ${\bar H}/({\sqrt
q}\beta) \propto -p=\sqrt{\frac{\lambda}{l-\frac{1}{3}}}$ from
Eq.\eqref{constraint_included}, and the lapse can then be determined
from the relation $K=(1-3l)p$ as
$N=\frac{\partial_ta}{a(t)\sqrt{(l-\frac{1}{3})\lambda}}$. This is a specialization of the formula \eqref{lapse}.  The explicit
form of the metric is then
\begin{eqnarray}
ds^2&=&-\frac{3(\partial_ta)^2}{\lambda
a^2(3l-1)}dt^2+a^2\left[\frac{dr^2}{1-kr^2}+r^2\left(d\theta^2+\sin^2\theta
d\phi^2\right)\right]\\
&=& -dt'^2+ a(t')^2\left[\frac{dr^2}{1-kr^2}+r^2\left(d\theta^2+\sin^2\theta d\phi^2\right)\right],
\end{eqnarray}
which is cast in the usual the Robertson-Walker form after reparametrizing the metric by identifying $dt' :=\frac{\sqrt{3}(\partial_ta)}{a\sqrt{\lambda(3l-1)}}dt =\frac{\sqrt{3}d\ln a}{\sqrt{\lambda(3l-1)}}$ wherein $\lambda =
2\Lambda_{eff} -\frac{6k}{a^2}+\frac{9k^2}{2\Lambda_{eff}a^4}$ depends on $a$. This can be integrated to yield
the time dependence of $a$ in terms of $t'$ as \begin{equation}a^2(t')=\frac{1}{2\Lambda_{eff}}
 \left[3k+e^{2(t'- t'_0)\sqrt{\frac{2}{3}(3l-1)\Lambda_{eff}}}\right].\end{equation}
At large values of $(t'-t'_0){\sqrt\Lambda_{eff}}$ the resultant metric expands exponentially regardless of $k$, and with $l=\frac{1}{2}$ (as in Einstein's theory) it then yields the usual de Sitter expansion with $a(t') \propto e^{\sqrt{\frac{\Lambda_{eff}}{3}}(t'-t'_0)}$ .

 \section{Motion of test particle}
 The motion of a test point particle of mass $m_0$  described canonically by $(x^i_{\rm P}, P_i)$ can be derived from the particle Hamiltonian \begin{align}H_{\rm P} =\int [N\sqrt{q^{ij}P_iP_j + m^2_0}- N^iP_i]\delta({\vec x}-{\vec x}_{\rm P})d^3x=
 N({\vec x}_{\rm P},t)\sqrt{q^{ij}({\vec x}_{\rm P},t)P_iP_j + m^2_0}- N^i({\vec x}_{\rm P}, t)P_i. \label{Ham}
 \end{align}
The canonical equation $\frac{dx^i_{\rm P}}{dt} = \{x^i_{\rm P}, H_{\rm P}\}_{\rm P.B.}$ relates the velocity and momentum by
$ \frac{dx^i_{\rm P}}{dt}+ N^i= \frac{Nq^{ij}P_j}{\sqrt{q^{ij}P_iP_j + m^2_0}}$. Inverting for $P_i$ in terms of $\frac{dx^i_{\rm P}}{dt}+ N^i$ results in the action,
\begin{align}\label{Lag}
S = \int [P_i\frac{dx^i_{\rm P}}{dt} - H_{\rm P}] dt =-m_0 \int dt\sqrt{ N^2 - q_{ij}(\frac{dx^i_{\rm P}}{dt}+ N^i)(\frac{dx^j_{\rm P}}{dt}+ N^j)}= -m_0\int \sqrt{-g_{\mu\nu}({\bf x}_{\rm P}, t)dx^\mu dx^\nu};
\end{align}
which is just the usual proper time action, on identifying the ADM metric $ds^2 =g_{\mu\nu}dx^\mu dx^\nu =-N^2 dt^2+ q_{ij}(dx^i+ N^idt)(dx^j+ N^jdt)$. Conversely, starting with the Lagrangian in the final step of \eqref{Lag}, the Hamiltonian of \eqref{Ham} is obtained. It follows that the particle will obey the geodesic equation
 in the background ADM metric. The derivation is insensitive to the particular form of the background lapse function and holds, in particular, for $N =\frac{\sqrt{q}(\partial_t \ln q^{1/3} -\frac{2}{3}\nabla_iN^i)}{4\beta\kappa {\bar H}}$ in the intrinsic time formulation.

The particle Hamiltonian of \eqref{Ham} is motivated by the fact that in the presence of the particle, the total Hamiltonian constraint, $H_{\rm T}= H_{\rm {\rm pure\,\, GR}} + \sqrt{q}E_{\rm P}=0$, is equivalently
\begin{eqnarray}
0=H_T&=& \frac{2\kappa}{\sqrt{q}}(-\beta{\tilde\pi} + {\bar H})(\beta{\tilde\pi} + {\bar H}) + E_{\rm P}\nonumber\\
&=& \frac{2\kappa}{\sqrt{q}}\left[ \left(-\beta{\tilde\pi} + \sqrt{{\bar H}^2 + \frac{\sqrt{q}E_{\rm P}}{2\kappa}}\right)\left(\beta{\tilde\pi} + \sqrt{{\bar H}^2 + \frac{\sqrt{q}E_{\rm P}}{2\kappa}}\right)\right].
\end{eqnarray}
This implies the Hamiltonian for evolution w.r.t. ADM coordinate time $t$ is (see, for instance, Ref.\cite{Soo_Hoi-Lai, Niall} and also the appendix)
\begin{eqnarray}
H_{\rm ADM} &=& \int\, d^3x \frac{(\partial_t \ln q^{1/3} -\frac{2}{3}\nabla_iN^i)}{\beta}\sqrt{{\bar H}^2 +\frac{\sqrt{q}E_{\rm P}}{2\kappa}} + N^iH_i\\
&=& \int\, d^3x \left(\partial_t \ln q^{1/3} -\frac{2}{3}\nabla_iN^i\right)\frac{\bar H}{\beta}\left(1+ \frac{\sqrt{q}E_{\rm P}}{4\kappa{\bar H}^2} + ...\right) + N^iH_i \label{21} \\
&\approx&\int\, d^3x \left(\partial_t \ln q^{1/3} -\frac{2}{3}\nabla_iN^i\right)\frac{\bar H}{\beta}+
NE_{\rm P}  + N^iH_i,
\end{eqnarray}
with $N=\frac{\sqrt{q}(\partial_t \ln q^{1/3} -\frac{2}{3}\nabla_iN^i)}{4\beta\kappa{\bar H}}$ being the lapse function of the background geometry, and retaining the expansion in \eqref{21} only to first-order in $\frac{\sqrt{q}E_{\rm P}}{\kappa{\bar H}^2}$ (this ratio compares of the particle's energy to the Hamiltonian density of the rest of the universe).
The EOM of the particle are then,
\begin{eqnarray}
\frac{dx^i_{\rm P}}{dt}&=&\{x^i_{\rm P}, H_{\rm ADM}\}_{\rm P.B.}\approx \{x^i_{\rm P}, (\int d^3x NE_{\rm P}) - N^iP_i\}_{\rm P.B.}= \{x^i_{\rm P}, H_{\rm P}\}_{\rm P.B.},\nonumber\\
\frac{dP_i}{dt}&=& \{P_i, H_{\rm ADM}\}_{\rm P.B.}\approx  \{P_i, H_{\rm P}\}_{\rm P.B.};
\end{eqnarray}
with $E_{\rm P} = \sqrt{q^{ij}P_iP_j + m^2_0}\,\delta({\vec x}-{\vec x}_{\rm P})$ and, consistently, $H_{\rm P}$ as in \eqref{Ham}.

\subsection{Geodesic equation, perihelion shift and bending of light for Painleve-Gullstrand metric}

It was already demonstrated that a test particle will obey the geodesic equation. We now analyze this motion
in Schwarzschild-de Sitter spacetime expressed in constant-curvature PG form (for simplicity we use the notation $t:= t_{PG}$ in this subsection),
\begin{equation*}
ds^2=-Bdt^2+\left[\left(\frac{dR}{\sqrt{B}}+dt\sqrt{A}\right)^2+R^2d\Omega^2\right];\qquad
A=\frac{2GM}{Rc^2}+\frac{1}{6}\left(2\Lambda_{eff}-\frac{6k}{a^2}
+\frac{9k^2}{2\Lambda_{eff} a^4}\right)R^2, B=1-k\frac{R^2}{a^2}.
\end{equation*}
With $x^\mu_{\rm P} =(t, R, \theta, \phi)$,  the geodesic equation $\frac{d^2x^\mu}{dp^2} + \Gamma^\mu_{\nu\alpha}\frac{dx^\nu}{dp}\frac{dx^\alpha}{dp}=0$  is equivalent to
 \begin{equation}\left(\begin{array}{c}
-\frac{R\sqrt{A}\dot{\theta}^2}{\sqrt{B}}-\frac{R\sqrt{A}\sin^2\theta\dot{\phi}^2}{\sqrt{B}}-\frac{\dot{R}^2A'}{2\sqrt{A}B^{3/2}}
-\frac{\sqrt{A}\dot{t}^2(A'-B')}{2\sqrt{B}}-\frac{\dot{t}\dot{R}(A'-B')}{B}+\ddot{t}\cr\cr
R(A-B)\dot{\theta}^2+R(A-B)\sin^2\theta\dot{\phi}^2+\frac{1}{2}(A-B)(A'-B')\dot{t}^2+
\frac{\sqrt{A}\dot{t}\dot{R}(A'-B')}{\sqrt{B}}+\frac{\dot{R}^2(A'-B')}{2B}+\ddot{R}\cr\cr
\frac{2\dot{R}\dot{\theta}}{R}-\sin\theta\cos\theta\dot{\phi}^2+\ddot{\theta}\cr\cr
\frac{2\dot{R}\dot{\phi}}{R}+2\cot\theta\dot{\theta}\dot{\phi}+\ddot{\phi}\end{array}\right)=0;\end{equation}\\
wherein derivatives w.r.t. $p$ and $R$ are denoted by $\dot{}$ and $'$.
Choosing the initial motion to lie in the equatorial plane ($\theta=\frac{\pi}{2}, \dot{\theta}=0$) implies $\ddot{\theta}=0$ as well.
Furthermore, $\frac{2\dot{R}\dot{\phi}}{R}+\ddot{\phi}=\frac{1}{R^2}\frac{d}{dp}\left(R^2\dot{\phi}\right)=0$ leads to conservation of $R^2\dot{\phi}=:L^2$. On geodesics, there is also constancy of
$g_{\mu\nu}\dot{x}^\mu\dot{x}^\nu=(A-B)\dot{t}^2+\frac{2\sqrt{A}\dot{t}\dot{R}}{\sqrt{B}}+\frac{\dot{R}^2}
{B}+R^2\dot{\phi}^2=-c^2\left(\frac{d\tau}{dp}\right)^2=:-E$. For time-like geodesics, we may use $dp= cd\tau$ and
$\dot{x}^\mu=\frac{dx^\mu}{cd\tau}$ , whereas $E=0$ for null geodesics. In general,
\begin{equation}
(A-B)\dot{t}^2+\frac{2\sqrt{A}\dot{t}\dot{R}}{\sqrt{B}}+\frac{\dot{R}^2}{B}+R^2\dot{\phi}^2=-\Delta\mbox{,}
\quad\;\;
\Delta=\left\{\begin{array}{l}0\;\;\mbox{null geodesic}\cr 1\;\;\mbox{time-like}.\end{array}\right.
\end{equation}
Consequently $ \dot{t}=\frac{-R^2\dot{R}\sqrt{AB}\pm\sqrt{R^4\dot{R}^2B^2-L^2R^2AB^2+L^2R^2B^3-\Delta
R^4B^2(A-B)}}{R^2B(A-B)}$.
Substituting this into the geodesic equation for the radial coordinate results in
$\frac{2L^2(A-B)-R(L^2+\Delta R^2)(A'-B')}{2R^3}+\ddot{R}=0$; so the effects are dependent only upon
$A-B=-\left(1-\frac{2GM}{Rc^2}-\frac{\Lambda_{eff}
R^2}{3}-\frac{3k^2R^2}{4\Lambda_{eff} a^4}\right)$.  The remaining dynamical equation is thus
\begin{align}
\ddot{R}+\frac{3GL^2M}{R^4c^2}-\frac{L^2}{R^3}+\Delta
\left(\frac{GM}{R^2c^2}-\frac{R\Lambda_{eff}}{3}-\frac{3k^2R}{4a^4\Lambda_{eff}}\right)=0.
\end{align}
The trajectories $R(\phi)$  (with $\frac{dR}{dp}=\frac{d\phi}{dp}\frac{dR}{d\phi}=\frac{L}{R^2}\frac{dR}{d\phi}$; $u:=\frac{1}{R}, \frac{dR}{dp}=-\frac{1}{u^2}\frac{L}{R^2}\frac{du}{d\phi}=-L\frac{du}{d\phi},
\frac{d^2R}{dp^2}=-L(Lu^2)(\frac{d^2R}{d\phi^2})=-L^2u^2(\frac{d^2R}{d\phi^2})),$
are thus governed by
$-L^2u^2\left(\frac{d^2u}{d\phi^2}\right)+\frac{3GML^2u^4}{c^2}-L^2u^3+\Delta\left(\frac{GMu^2}{c^2}-
\frac{\Lambda_{eff}}{3u}-\frac{3k^2}{4a^4\Lambda_{eff} u}\right)=0$,  or equivalently,
\begin{align}\label{precession}
\frac{d^2u}{d\phi^2}+u-\frac{3GMu^2}{c^2}=\Delta\left(\frac{GM}{L^2c^2}-\frac{\Lambda_{eff}}{3L^2u^3}\left(1+
\frac{9k^2}{4a^4\Lambda^2_{eff}}\right)\right).
\end{align}
 This is precisely the {\it same equation as in Einstein's theory} with cosmological constant  provided, as motivated in the earlier discussion, $\frac{k^2}{a^4\Lambda^2_{eff}}\mapsto 0$.
 Thus the motion of a test particle in constant curvature PG exact solution of intrinsic time geometrodynamics is in complete agreement with the predictions of Einstein's GR.
\subsection{Comparison with non-constant curvature solutions in Horava Gravity}
There exists other solutions with zero shift in Horava gravity\cite{Pope, Yavartanoo, Ghodsi, Aliev}.
The solution of Ref.\cite{Pope} is
$ds^2=-N^2dt^2+\frac{dr^2}{f}+r^2\left(d\theta^2+\sin^2\theta
d\phi^2\right)$, wherein $f=1-\frac{2}{3}\Lambda_{eff}r^2-\alpha
r^{\frac{1}{2}+\frac{9(1-2l)}{2\left(1+\sqrt{6l-2}\right)^2}}$. Even when
$l=\frac{1}{2}$, the solution with
$f=1-\frac{2}{3}\Lambda_{eff}r^2-\alpha\sqrt{r}$ deviates from the
 Schwarzschild form and it is not of constant 3-curvature.
Setting the deformation parameter to $l=\frac{1}{2}$ for comparison with Einstein's theory yields
\begin{equation}
ds^2=-N^2 dt^2+\frac{dr^2}{f}+r^2\left(d\theta^2+\sin^2\theta
d\phi^2\right); \hspace{1 cm} N^2=f=1-\frac{2}{3}\Lambda
r^2+\alpha\sqrt{r}.
\end{equation}
It can be shown that the corresponding geodesic equation takes the form
\begin{equation}\label{lousy}
\frac{d^2u}{d\phi^2}+u+\frac{3\alpha\sqrt{u}}{4}=\Delta\left(\frac{\alpha}{4L^2u^{3/2}}-\frac{2\Lambda}{3L^2u^3}\right),
\hspace{1 cm}
\Delta=\left\{\begin{array}{l}0\;\;\mbox{null geodesic}\cr 1\;\;\mbox{time-like},\end{array}\right.
\end{equation}
which differs from Eq.\eqref{precession}, and thus from the predictions of Einstein's theory.
Solving for the geodesics numerically with fixed $L$ and initial conditions $u(0)$,
$u'(0)=0$  yields the results in Figs.1-4 which provide stark graphical comparisons of the predictions of
 \eqref{precession} and \eqref{lousy}. Unlike  \eqref{precession} in intrinsic time geometrodynamics, \eqref{lousy} which has a very different dependence on $u$ fails to produce the `normal precession of perihelion  behavior' in Einstein's theory for time-like bound geodesics
 even when the parameter in the solution $\alpha$ is varied over a wide range.

It should be pointed out that the examples discussed previously were, for concrete explicit comparisons, all based upon the form of  $\bar H$ in (4) and (5) wherein, for constant spatial curvature slicings, the departure from Einstein's theory consists of only a term proportional to ${\cal R}^2$. For the explicit form of $\bar H$ advocated in Ref.[3], departure from the potential of Einstein's theory with cosmological constant is $q(c'{\bar{\cal R}}_{ij}+ g'C_{ij})(c'{\bar{\cal R}}^{ij}+ g'C^{ij})$, which implies that the results for constant spatial curvature solutions will agree completely with Einstein's theory.

\begin{figure}
\centering
\begin{minipage}{.4\textwidth}
\centering
\includegraphics[width=0.8\textwidth]{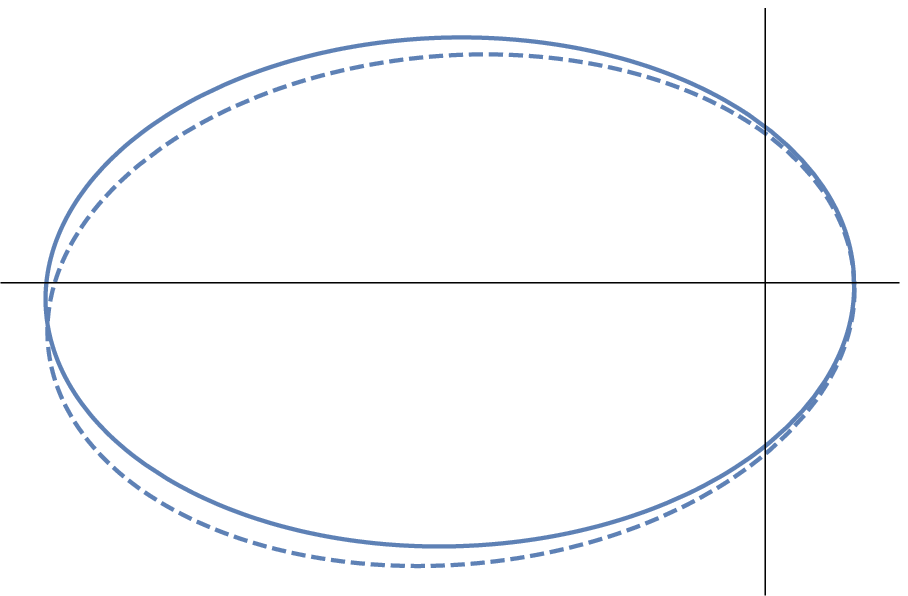}
\caption{}
\end{minipage}
\begin{minipage}{.4\textwidth}
\centering
\includegraphics[width=0.8\textwidth]{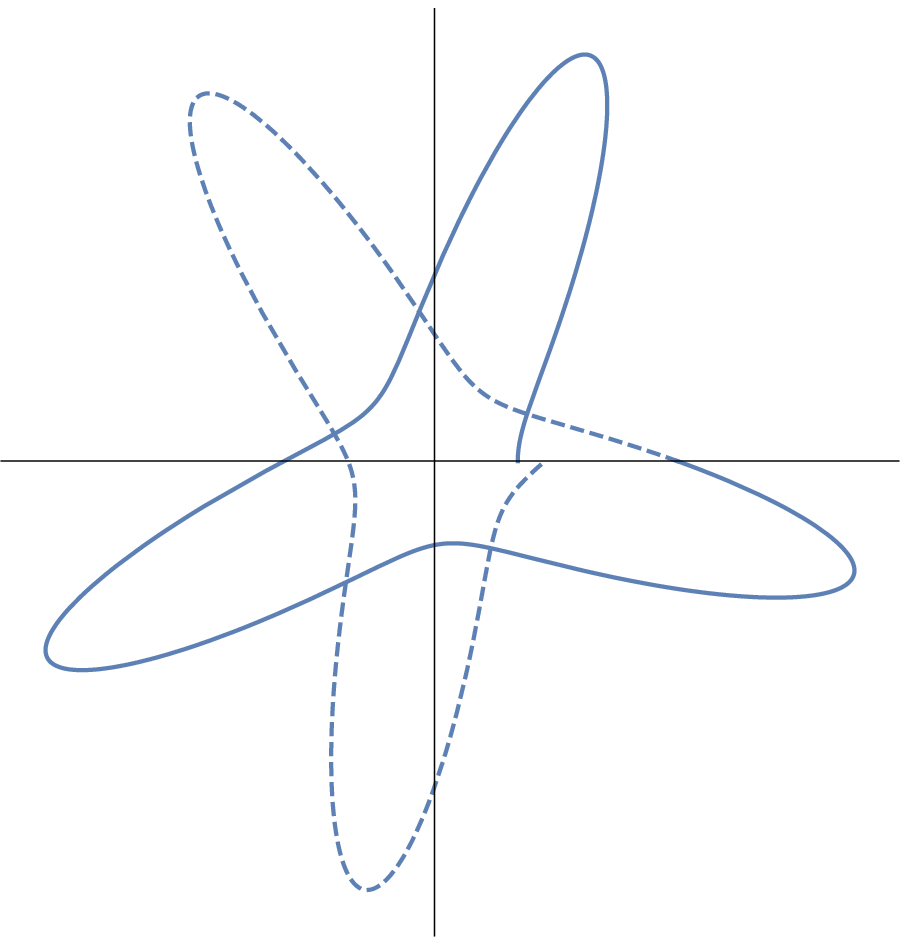}
\caption{}
\end{minipage}
\begin{minipage}{.4\textwidth}
\centering
\includegraphics[width=0.8\textwidth]{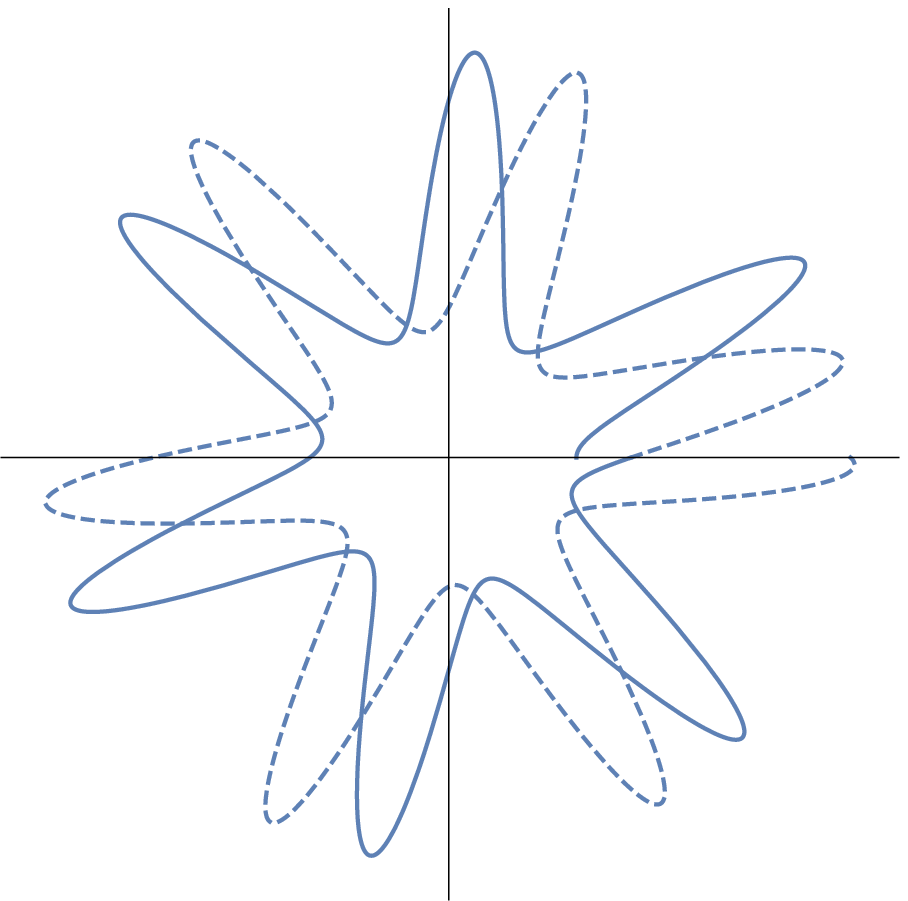}
\caption{}
\end{minipage}
\begin{minipage}{.4\textwidth}
\centering
\includegraphics[width=0.8\textwidth]{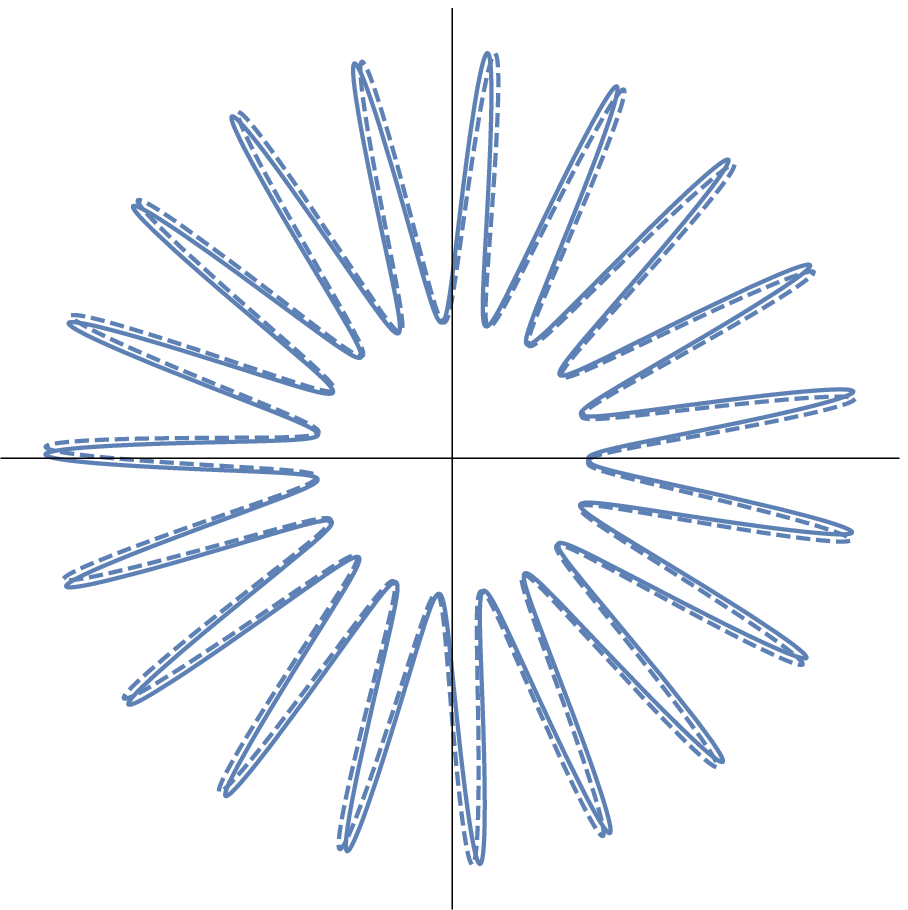}
\caption{}
\end{minipage}
\caption*{Fig. 1 shows the usual precession of perihelion behavior predicted by Eq. (32) of intrinsic time geometrodynamics; whereas Figs. 2-4 are the starkly different predictions of  Eq. (34) as the parameter $\alpha$ is varied over a wide range.
All figures have the same $L$ and initial conditions $u(0), u'(0)$; and the orbits are shown for `two revolutions' i.e. $0 \leq \phi \leq 4\pi$,  with solid lines for $0 \leq \phi < 2\pi$  and dashed lines for $2\pi \leq \phi \leq 4\pi$.}
\end{figure}

\newpage
\section*{Appendix}

The Hamiltonian constraint of General Relativity can be expressed as
\begin{eqnarray}
0\cong \frac{\sqrt{q}}{2\kappa} H &=& {G}_{ijkl}{\tilde\pi}^{ij}{\tilde\pi}^{kl} -\beta^2{\tilde\pi}^2 + V(q_{ij})\nonumber\\
&=&{\bar G}_{ijkl}{\bar\pi}^{ij}{\bar\pi}^{kl} -\beta^2{\tilde\pi}^2 + V(q_{ij})  \nonumber\\
&=:& \bar H^2 -\beta^2{\tilde\pi}^2 \nonumber\\
&=&({\bar H}-\beta{\tilde\pi})({\bar H}+\beta{\tilde\pi}). \nonumber\\
\end{eqnarray}
This constrains $ \bar H=\pm\beta{\tilde\pi}$, wherein $\bar {H}(\bar{\pi}^{ij}, \bar {q}_{ij}, q)  := \sqrt{ \bar{G}_{ijkl}\bar{\pi}^{ij}\bar{\pi}^{kl} +  V(\bar {q}_{ij}, q) }$. For Einstein's General Relativity,  $l = 1$ and $V({q}_{ij}) =- \frac{q}{(2\kappa)^2}[R - 2\Lambda_{\it{eff}} ]$, the constraints form a first class algebra, and the lapse function $N$ is {\it a priori} arbitrary.
But $N$ is fixed {\it a posteriori} by the EOM and constraints.  In particular, the lapse function is related to $\partial_t\ln q$ and the Hamiltonian density ${\bar H}$ through
\begin{eqnarray}
\frac{\partial{\qt}(x)}{\partial t} &=& \left\{ \qt(x), \int N(y)H(y)d^3y \right\}_{P.B.}=\int N(y)\left\{ \qt(x), H(y)\right\}_{P.B.} d^3y\\
&=& -\frac{4{N}(x)}{\sqrt q}\kappa\beta^2\tilde\pi(x)\\
&=& \mp 4\kappa\beta\underset{\sim}{N}(x){\bar H}(x),
\end{eqnarray}
wherein the Poisson bracket $\{ \ln q^{\frac{1}{3}}(x), {\tilde\pi}(y)\}_{P.B.} =\delta(x-y)$ has been used to reach the intermediate step (37), and the Hamiltonian constraint in the last step (even though $N$ was a priori arbitrary).
Remarkably, adopting the non-vanishing Hamiltonian,  $H_{\rm ADM}  := \frac{1}{\beta}\int \frac{\partial\ln q^{\frac{1}{3}}(x)}{\partial t}{\bar H}(x) d^3x$  which generates evolution w.r.t. ADM coordinate time $t$, and {\it dispensing with} the Hamiltonian constraint will yield classical physics equivalent to Einstein's GR since
\begin{eqnarray}
{\dot{\bar q}}_{ij}(x) &=& \left\{ {\bar q}_{ij}(x),  \frac{1}{\beta}\int \frac{\partial\qt(y)}{\partial t}{\bar H}(y)d^3y\right\}_{P.B.}\\
&=&\int \frac{1}{\beta}\frac{\partial\qt(y)}{\partial t}\left\{ {\bar q}_{ij}(x), {\bar H}(y)\right\}_{P.B.} d^3y\\
&=& \int \mp\underset{\sim}{N}(y)4\kappa{\bar H}(y)\left\{ {\bar q}_{ij}(x), {\bar H}(y)\right\}_{P.B.} d^3y\\
&=&\int \mp\underset{\sim}{N}(y)2\kappa\left\{ {\bar q}_{ij}(x), {\bar H}^2(y)\right\}_{P.B.} d^3y\\
&=&\int \mp{N}(y)\left\{ {\bar q}_{ij}(x), H(y)\right\}_{P.B.} d^3y.
\end{eqnarray}
This is equivalent to the evolution generated through $\int {\mp N}(y)\left\{ {\bar q}_{ij}(x), H(y)\right\}_{P.B.} d^3y$ with the a posteriori relation (38) between $N, \partial_t\ln q$ and ${\bar H}$.
The $\mp$ sign accompanying $N$ does not matter because the resultant classical ADM four-metric depends only on the square of $N$. Adding $\int N^i(y)H_i(y) d^3y$ to the Hamiltonian merely leads to modification of the EOM by Lie derivatives
of the variables w.r.t. $N^i$, with the resultant relation (38) generalized to the form displayed earlier in (3). The EOM for $\bar\pi^{ij}$ can be similarly demonstrated. In fact (3) ensures the
Hamiltonian constraint is satisfied classically in the form $\frac{K^2}{9} = \frac{4\kappa^2\beta^2}{q}{\bar H}^2$. Thus for the true d.o.f., $(\bar q_{ij}, {\bar\pi}^{ij})$, the non-trivial Hamiltonian $H_{ADM}$ equivalently captures the physical content and EOM of Einstein's theory. This framework however allows the potential $V$ to depart from that of Einstein's theory without leading to inconsistencies in the constraint algebra.

While $q$ is a tensor density, the multi-fingered intrinsic time interval,  $\delta\qt =\frac{1}{3}\frac{\delta q}{q}= \frac{q^{ij}}{3}\delta q_{ij}$, is a scalar entity.
Hodge decomposition for any compact Riemannian manifold without boundary yields
$\delta \ln q^{\frac{1}{3}} =\delta{T}+\nabla_i\delta{Y}^i$, wherein the gauge-invariant part of $\delta\qt$ is $\delta T =\frac{2}{3}\delta \ln V_{\rm spatial}$ which is proportional to the 3dDI logarithmic change in the spatial volume.
This can be seen from $\int (\delta\qt) \sqrt{q} d^3x = \int (\delta{T}+\nabla_i\delta{Y}^i) \sqrt{q} d^3x =(\delta{T})\int \sqrt{q} d^3x =(\delta T)V_{\rm spatial}$ (which implies that $\delta T$ is the average value of $\delta\qt(x)$ over the spatial volume),
 and also $\int (\delta\qt )\sqrt{q} d^3x =\frac{1}{3}\int \frac{(\delta{\sqrt{q}}^2)}{{\sqrt q}^2} \sqrt{q}d^3x =\frac{2}{3}\delta(\int \sqrt{q}d^3x)=\frac{2}{3}\delta V_{\rm spatial}$.
 Instead of discussing the evolution of quantum states w.r.t.  gauge-dependent multi-fingered time $\qt(x)$, it is eminently more meaningful to ask how the wave function of the universe, $\Psi$, changes w.r.t to the 3dDI global variable $T$.
With $T$ displacing the physically less concrete ADM coordinate time $t$ in $H_{ADM}$, this dynamics is determined  by the Schrodinger equation, $i\hbar \frac{\delta \Psi}{\delta T} = {H}_{\rm Phys}\Psi$, wherein $H_{\rm Phys} := \int \frac{\bar H}{\beta} d^3x$ is the physical Hamiltonian generating evolution in global intrinsic time.  The formalism is applicable to the full theory of quantum gravity\cite{ITQG, NCR}, and assumptions of mini-superpsace models have not been invoked.
$\beta >0$ is required for the Hamiltonian to be bounded from below, and global intrinsic time increases monotonically with our ever expanding universe.
 \\
 \newpage
\section*{Acknowledgments}
This work was supported in part by the Ministry of Science and Technology (R.O.C.) under Grant Nos.
NSC101-2112-M-006 -007-MY3 and MOST104-2112-M-006-003. We would like to thank Eyo Eyo Ita III and Hoi-Lai Yu for beneficial discussions during the course of this work.
\\

\end{document}